\documentclass[12pt,a4paper]{article}
\usepackage{graphicx,psfrag,epsf,color}
\usepackage[dvipsnames]{xcolor}
\usepackage{amsmath,amssymb,amsfonts,bm}
\usepackage[
    bookmarksnumbered=true,
    urlcolor=blue,
    linkbordercolor=red,
    citebordercolor=green,
    bookmarksopen=true
    ]{hyperref}
\usepackage{geometry}
\usepackage{float}
\usepackage{mathrsfs}  
\usepackage{authblk}
\usepackage{array}

\usepackage[labelfont=bf]{caption}

\usepackage{cite}
\newcommand{\tr}[1]{\langle #1 \rangle}
\newcommand{\com}[2]{[#1,#2]}
\newcommand{\anticom}[2]{\{#1,#2\}}

\newcommand{\T}{\bm{T}}
\newcommand{\Id}{\bm{1}}
\renewcommand{\S}{\bm{S}}
\newcommand{\Oj}{\bm{O}^{(j)}}
\renewcommand{\H}{\bm{\mathcal{H}}}
\newcommand{\W}{\bm{\mathcal{W}}}
\newcommand{\VG}{\bm{V}^G}
\newcommand{\GammaH}{\bm{\Gamma}^H}
\newcommand{\GammaS}{\bm{\Gamma}^S}
\newcommand{\Gammac}{\bm{\Gamma}^c}
\newcommand{\GammaC}{\bm{\Gamma}^C}
\newcommand{\V}{\bm{V}}
\newcommand{\R}{\bm{R}}
\newcommand{\GammaBar}{\overline{\bm{\Gamma}}}
\newcommand{\Rc}{\bm{R}^c}
\newcommand{\RBar}{\overline{\bm{R}}}
\newcommand{\Vc}{\bm{V}^c}
\newcommand{\VBar}{\overline{\bm{V}}}

\numberwithin{equation}{section}

\renewcommand{\title}{Glauber Phases in Non-Global LHC Observables: \texorpdfstring{\\[1ex]}{}Resummation for Quark-Initiated Processes}

\begin{document}
\allowdisplaybreaks

\begin{titlepage}

\begin{flushright}
{\small
MITP-23-010\\
{July 20, 2023}
}
\end{flushright}

\vskip0.8cm
\pdfbookmark[0]{\title}{title}
\begin{center}
{\Large \bf\boldmath \title}
\end{center}

\vspace{0.5cm}
\begin{center}
    \textsc{Philipp~B\"oer,$^a$ Matthias Neubert$^{a,b}$ and Michel Stillger$^a$} \\[6mm]
    
    \textsl{${}^a$PRISMA$^+$ Cluster of Excellence \& Mainz Institute for Theoretical Physics\\
    Johannes Gutenberg University, 55099 Mainz, Germany\\[0.3cm]
    ${}^b$Department of Physics \& LEPP, Cornell University, Ithaca, NY 14853, U.S.A.}
\end{center}

\vspace{0.6cm}
\pdfbookmark[1]{Abstract}{abstract}
\begin{abstract}
\vskip0.2cm\noindent
It has been known for many years that jet cross sections at hadron colliders exhibit double-logarithmic corrections starting at four-loop order, arising from two soft Glauber-gluon interactions between the two colliding partons. The resummation of these ``super-leading logarithms'' has been achieved only recently by means of a renormalization-group treatment in soft-collinear effective theory. We generalize this result and, within the same framework and for quark-initiated processes, resum the double logarithms arising in the presence of an arbitrary number of Glauber-gluon exchanges. For typical choices of parameters, the higher-order Glauber terms give rise to corrections which are expected to be numerically of the same magnitude as the super-leading logarithms. However, we find that the Glauber series for jet cross sections is dominated by the two-Glauber contribution, if the colliding partons are quarks or anti-quarks.
\end{abstract}

\vfill\noindent\rule{0.4\columnwidth}{0.4pt}\\
\hspace*{2ex} {\small \textit{E-mail:} \href{mailto:pboeer@uni-mainz.de}{pboeer@uni-mainz.de}, \href{mailto:matthias.neubert@uni-mainz.de}{matthias.neubert@uni-mainz.de}, \href{mailto:m.stillger@uni-mainz.de}{m.stillger@uni-mainz.de}}

\end{titlepage}

\section{Introduction}
\label{sec:intro}

In April 2022, the LHC has continued its high-luminosity run, 12 years after the first protons were collided. In the absence of any direct discoveries of new physics, the question poses itself which strategy one should take to fully exploit the discovery potential of the high-luminosity LHC. In recent years, it has become clear that precision may be the key to discovery. Indirect signals of new physics might already be hiding in the LHC data, but we are limited in our ability to discover them due to present theoretical uncertainties. In searches for new phenomena, the Standard Model background processes must be controlled with highest possible accuracy. Thus, there is a need to significantly improve our ability to calculate the cross sections for important LHC processes, both in the Standard Model and in extensions featuring new particles. 

Scattering processes in which jets -- highly collimated sets of energetic particles -- are produced are among the most important observables studied in high-energy collisions, because they closely mirror the underlying hard-scattering event. They are thus well suited to study short-distance physics and play an important role in the search for new phenomena. However, the rates for jet production at hadron colliders are also among the most complicated observables to calculate theoretically. In non-global observables, such as exclusive jet cross sections~\cite{Sterman:1977wj}, one vetoes radiation with energies or transverse momenta above a low scale $Q_0$ in certain phase-space regions. A simple example are gap-between-jets cross sections with a veto on radiation outside the jets, whereas the particles inside the jets can carry large energies of order the hard scale $Q\equiv\sqrt{\hat s}$ set by the partonic center-of-mass energy. The coefficients of the perturbative series for such observables are enhanced by large logarithms $L=\ln(Q/Q_0)$, which include the so-called non-global logarithms arising from secondary emissions off the original hard partons~\cite{Dasgupta:2001sh}. In higher orders of perturbation theory, the logarithmically-enhanced contributions in these observables have a very intricate structure. At hadron colliders, subtle soft interactions between the two initial-state partons in the collision, mediated by the exchange of so-called Glauber gluons, lead to a breakdown of color coherence, a property implicitly assumed in all traditional calculations of jet cross sections. Starting at four-loop order, this generates a series of double-logarithmic corrections known as ``super-leading logarithms'' (SLLs)~\cite{Forshaw:2006fk,Forshaw:2008cq,Keates:2009dn}, which are related to a non-cancellation of initial-state collinear singularities~\cite{Catani:2011st,Forshaw:2012bi,Schwartz:2017nmr}. Since they arise from gluon exchange between initial-state partons, the SLLs appear only for observables at hadron colliders. 

Quite recently, their all-order resummation has been accomplished for generic $2\to M$ hard-scattering processes~\cite{Becher:2021zkk,Becher:2023mtx} using a novel type of factorization theorem~\cite{Becher:2015hka,Becher:2016mmh,Balsiger:2018ezi} derived in soft-collinear effective theory (SCET)~\cite{Bauer:2000yr,Bauer:2001yt,Beneke:2002ph,Beneke:2002ni} and solving the associated renormalization-group (RG) evolution equations. One finds that the SLL series is of the form
\begin{align}
   \sigma^{\rm SLL} 
   &\sim \frac{1}{N_c^2}\,\frac{N_c\,\alpha_s}{\pi}\,L \left| \frac{N_c\,\alpha_s}{\pi}\,i\pi\,L \right|^2
    \sum_{n=0}^\infty c_{1,n} \left( \frac{N_c\,\alpha_s}{\pi}\,L^2 \right)^n \nonumber\\
   &= \frac{\alpha_s\,L}{\pi\,N_c} \left( \frac{N_c\,\alpha_s}{\pi}\,\pi^2 \right)
    \sum_{n=0}^\infty\,c_{1,n} \left( \frac{N_c\,\alpha_s}{\pi}\,L^2 \right)^{n+1} ,
\end{align}
where the complex phase $i\pi$ arises from the imaginary part of \mbox{$\ln[(-Q^2-i0)/Q_0^2]=2L-i\pi$}, and the prefactor $1/N_c^2$ in the first line indicates that SLLs arise at subleading order in the large-$N_c$ expansion.
For typical values of parameters, e.g.\ $Q=1$\,TeV, $Q_0=40$\,GeV and $\alpha_s=\alpha_s(\sqrt{Q Q_0})$, one finds that the quantities
\begin{equation}
\label{eq_wwpi}
   w = \frac{N_c\,\alpha_s}{\pi}\,L^2 \qquad \text{and} \qquad
   w_\pi = \frac{N_c\,\alpha_s}{\pi}\,\pi^2 
\end{equation}
are both approximately equal to 1. As a result, repeated insertions of pairs of Glauber exchanges lead to a more general series of super-leading terms, to which we refer as the ``Glauber series'' \cite{Becher:2023mtx}. Schematically, we obtain
\begin{align}
\label{eq_Glauberseries_generic}
   \sigma^{\rm SLL+G}
   \sim \frac{\alpha_s\,L}{\pi\,N_c}\,\sum_{\ell=1}^\infty\sum_{n=0}^\infty \,c_{\ell,n}\,w_\pi^\ell\,w^{n+\ell} \,.
\end{align}
For parameters such that $w\sim w_\pi=\mathcal{O}(1)$, one might expect the double sum to give effects which numerically can be as large as the one-loop prefactor. Treating both the real and imaginary parts of the large logarithm $\ln[(-Q^2)/Q_0^2]$ on the same footing, we see that the double-logarithmic behavior of the series starts at three-loop order ($\ell=1$ and $n=0$).

So far, very little is known about the higher-order structure of the Glauber series. Like the SLLs themselves, none of the terms in \eqref{eq_Glauberseries_generic} are included in conventional parton showers. To improve the treatment of color in a parton shower, an exponentiation of the Glauber operator was suggested in~\cite{Nagy:2019rwb}, resulting in small net effects. It is of importance to get an improved analytic understanding of the Glauber series and pin down if, and for which processes, the higher-order terms give rise to non-negligible contributions. In the present paper, we show how to resum the Glauber series to all orders in perturbation theory for scattering processes for which both initial-state partons transform in the fundamental (or anti-fundamental) representation of $SU(N_c)$. To do so, we derive an explicit expression for the coefficients $c_{\ell,n}$ in the Glauber series for generic values of $\ell$ and $n$, proving in particular that the series is alternating, i.e.\
\begin{align}
   c_{\ell,n}\propto(-1)^{n+\ell} \,.
\end{align}
Given the complexity of the color algebra involved in this problem, it may seem surprising that an analytic all-order expression for the expansion coefficients can be obtained. However, we show that the application of the respective operations in color space form a closed system, and that the relevant color traces can always be reduced to structures not more complicated than those arising in the calculation of the SLLs, at least for processes with quarks or anti-quarks in the initial state. We believe that our analytic results provide valuable insights benefitting the ongoing efforts to improve parton showers beyond the large-$N_c$ approximation, see e.g.~\cite{Nagy:2019pjp,DeAngelis:2020rvq,Hoche:2020pxj,Hamilton:2020rcu,Becher:2023vrh}.

The remainder of this article is organized as follows: In Section~\ref{sec:basics} we introduce some basic definitions and recapitulate the derivation of the series of SLLs for the gap-between-jets cross section. The generalization of the color traces to the case of the higher-order terms in the Glauber series (with $\ell\ge 2$) and the reduction of these traces to elementary building blocks is described in Section~\ref{sec_generalized_color_trace}. In Section~\ref{sec_nested_integrals} we combine this result with the nested scale integrals arising from an iterative solution of the RG equation to obtain an expression for the Glauber series coefficients $c_{\ell,n}$. A numerical estimate of the higher-order Glauber contributions to some simple partonic scattering processes is presented in Section~\ref{sec_pheno}, before we conclude in Section~\ref{sec_conclusions}.

\section{Basic definitions and super-leading logarithms}
\label{sec:basics}

Starting point of our analysis is the factorization formula for the gap-between-jets cross section for a $2 \to M$ wide-angle jet event at hadron colliders, which reads~\cite{Balsiger:2018ezi,Becher:2021zkk,Becher:2023mtx}
\begin{align} \label{eq_cross_section_general}
	\sigma_{2\to M}(Q_0) = \int\!dx_1 \! \int\!dx_2 \! \sum_{m=2+M}^{\infty} \tr{\H_m(\{\underline{n}\},s,x_1,x_2,\mu) \otimes \W_m(\{\underline{n}\},Q_0,x_1,x_2,\mu)} \,.
\end{align}
Here the hard functions $\H_m$ describe all possible partonic on-shell hard-scattering processes $1+2 \to 3 + \dots + m$ with appropriate kinematic constraints~\cite{Becher:2023mtx},
\begin{align} \label{eq_def_hard_function}
	\H_m &= \frac{1}{2x_1 x_2 s} \prod_{i=3}^m \int \frac{dE_i E_i^{d-3}}{\tilde{c}^\epsilon  (2\pi)^2} \, |\mathcal{M}_m(\{\underline{p}\})\rangle\langle\mathcal{M}_m(\{\underline{p}\})| \nonumber \\
	&\quad\times (2\pi)^d \, 2 \, \delta(\bar{n}_1\cdot p_{\text{tot}}-x_1\sqrt{s}) \, \delta(\bar{n}_2\cdot p_{\text{tot}}-x_2\sqrt{s}) \, \delta^{(d-2)}(p_{\text{tot}}^\perp) \, \Theta_{\text{hard}}(\{\underline{n}\}) \,,
\end{align}
whereas the low-energy matrix elements $\W_m$ describes the soft and collinear dynamics. Both the hard functions and the low-energy matrix elements depend on the directions $\{\underline{n}\} = \{n_1, \dots, n_m\}$ of the particles involved in the scattering. The symbol $\otimes$ denotes an integration over the directions $\{n_3,\dots,n_m\}$ of the
final-state partons. Throughout this paper, we use the color-space formalism~\cite{Catani:1996vz} and closely follow the notation of~\cite{Becher:2021zkk,Becher:2023mtx}, which we do not repeat here.

The scale dependence of the hard functions is gouverned by the RG equation
\begin{align} \label{eq:HmRGE}
    \frac{d}{d\ln\mu}\,\H_m (\{\underline{n}\},s,\mu) = - \sum_{l=2+M}^m \H_l(\{\underline{n}\},s,\mu) \star \GammaH_{lm}(\{\underline{n}\},s,\mu) \,,
\end{align}
where the $\star$ symbol indicates a Mellin convolution in the partonic momentum fractions $x_1$ and $x_2$. Here and below we omit the momentum-fraction variables for the initial-state partons in the hard functions and the anomalous dimension. This equation has the structure of an RG equation in the presence of operator mixing. However, solving this equation is a highly non-trivial task. The reason is that the anomalous-dimension matrix is an operator not only in the high-dimensional color space of the initial- and final-state particles, but also in the infinite space of particle multiplicities. The evolution equation shows that higher-multiplicity hard functions mix with lower-multiplicity functions under scale evolution. A formal solution of \eqref{eq:HmRGE} can be written in terms of a path-ordered exponential
\begin{align}
    \bm{U}(\{\underline{n}\},s,\mu_h,\mu) = \mathbf{P} \exp\left[\int_{\mu}^{\mu_h}\frac{d\mu'}{\mu'} \GammaH(\{\underline{n}\},s,\mu')\right] ,
\end{align}
which is defined by its series expansion (schematically, omitting irrelevant arguments as well as multiplicity indices)
\begin{align}
\label{eq_evolution_series}
    &\H(\mu_h) \star \bm{U}(\mu_h,\mu) \nonumber \\
    &= \H(\mu_h) + \int_\mu^{\mu_h}\!\frac{d\mu'}{\mu'}\,\H(\mu_h) \star \GammaH(\mu') 
     + \int_\mu^{\mu_h}\!\frac{d\mu'}{\mu'} \int_{\mu'}^{\mu_h}\!\frac{d\mu''}{\mu''}\,
     \H(\mu_h) \star \GammaH(\mu'') \star \GammaH(\mu') + \dots \,.
\end{align}
The one-loop anomalous dimension $\GammaH = \GammaC + \GammaS$ can be split into a purely collinear part $\GammaC$ (see \cite{Becher:2023mtx} for a detailed discussion of this quantity) as well as soft and soft-collinear terms $\GammaS$, for which the Mellin convolution is trivial, since soft partons can only take away an insignificant amount of momentum. We will thus omit the $\star$ symbol when writing products of soft and soft-collinear anomalous dimensions below. In the remainder of this article we will ignore $\GammaC$, as it is of subleading logarithmic order and does not contain Glauber phases. The soft part takes the form
\begin{align}
	\GammaS(\{\underline{n}\},s,\mu) = \frac{\alpha_s}{4\pi}
	\begin{pmatrix}
		\V_{2+M}^S & \R_{2+M\phantom{+1}}^S & 0 & 0 & \dots \\[0.2ex]
		0 & \V_{2+M+1}^S & \R_{2+M+1}^S & 0 & \dots \\[0.2ex]
		0 & 0 & \V_{2+M+2}^S & \R_{2+M+2}^S & \dots \\[0.2ex]
		0 & 0 & 0 & \V_{2+M+3}^S & \dots \\[0.2ex]
		\vdots & \vdots & \vdots & \vdots & \ddots
	\end{pmatrix} + \mathcal{O}(\alpha_s^2) \,.
\end{align}
The virtual contributions on the diagonal leave the number of partons unchanged, while real-radiation contributions on the first off-diagonal increase the parton multiplicity by one. Double-real emissions enter at $\mathcal{O}(\alpha_s^2)$ and would occupy the second off-diagonal~\cite{Becher:2021urs}, and so on. At $\mathcal{O}(\alpha_s)$ the entries can be further split into
\begin{align}
\label{eq_VSRS}
	\V_m^S &= \VBar_{\!\!m} + \VG + \sum_{i=1,2} \Vc_i \, \ln\frac{\mu^2}{\mu_h^2} \,,
	\nonumber\\
	\R_m^S &= \RBar_m + \sum_{i=1,2} \Rc_i \, \ln\frac{\mu^2}{\mu_h^2} \,,
\end{align}
where explicit expressions for individual pieces can be found in eq.~(7) in \cite{Becher:2021zkk}. Importantly, the operators $\VG$, $\Vc_i$ and $\Rc_i$ only involve the color generators of the initial-state partons ($i=1,2$), whereas the operators $\VBar_{\!\!m}$ and $\RBar_m$ involve color generators for all partons in the process. The most relevant terms for the purposes of this work are
\begin{align}  
   \VG &= - 2i\pi\,\gamma_0^{\rm cusp}\,\big( \bm{T}_{1,L}\cdot\bm{T}_{2,L} - \bm{T}_{1,R}\cdot\bm{T}_{2,R} \big) 
    \,, \notag\\
   \Vc_i &= \gamma_0^{\rm cusp}\,C_i\,\bm{1} \,, \notag\\
   \Rc_i &= - \gamma_0^{\rm cusp}\,\bm{T}_{i,L}\circ\bm{T}_{i,R}\,\delta(n_{k}-n_i) \,,
\end{align}
where color generators $\bm{T}_{i,L}$ ($\bm{T}_{i,R}$) act on the $i$-th parton in the amplitude (complex conjugate amplitude), $C_i$ denotes the eigenvalue of the quadratic Casimir operator for parton $i$, the symbol $\circ$ indicates that when a gluon is emitted by application of the operator $\Rc_i$ a new color space is created, and the $\delta$-function ensures that this gluon (with index $k$) is collinear with the direction $n_i$ of the parton off which it was radiated. These operators are proportional to the cusp anomalous dimension, for which we use the one-loop coefficient $\gamma_0^{\text{cusp}}=4$ in the following. For our analysis we combine the real and virtual parts as follows
\begin{align}
	\Gammac \equiv \sum_{i=1}^{2} \left(\Vc_i + \Rc_i\right) , \qquad
	\H_m \, \GammaBar \equiv \H_m \left(\VBar_{\!\!m} + \RBar_m\right) \,.
\end{align}

To sum logarithmically-enhanced quantum corrections we then evolve the hard functions from their characteristic high scale $\mu_h \sim Q=\sqrt{\hat{s}}$ down to the soft veto scale $\mu_s\sim Q_0$. To obtain the leading double logarithms (the SLLs) it is then sufficient to evaluate the low-energy matrix elements $\W_m(\mu_s)$ at lowest-order, i.e.\
\begin{align}
\label{eq_LOlowenergyME}
	\W_m(\{\underline{n}\},Q_0,x_1,x_2,\mu_s) = f_1(x_1,\mu_s) \, f_2(x_2,\mu_s) \, \Id + \mathcal{O}(\alpha_s)\,,
\end{align}
with $f_i(x_i)$ being the standard parton distribution functions. In the following we briefly summarize the relevant results from \cite{Becher:2021zkk,Becher:2023mtx} with a focus on quark-initiated processes. As a consequence of the three relations
\begin{align}
    [ \Gammac,\GammaBar ] = 0 \,, \qquad
    \big\langle \bm{\mathcal{H}}\,\Gammac\otimes\bm{1} \big\rangle = 0 \,, \qquad
    \big\langle \bm{\mathcal{H}}\,\VG\otimes\bm{1} \big\rangle = 0 \,,
\end{align}
where $\bm{\mathcal{H}}$ can be an arbitrary hard function, one finds that the color traces associated with the SLLs are
\begin{align} \label{eq_color_trace_SLL}
    C_{r_1,r_2} = \tr{\H_{2\to M}\left(\Gammac\right)^{r_1} \VG \left(\Gammac\right)^{r_2} \VG \, \GammaBar \otimes \Id} \,,
\end{align}
where $\H_{2\to M}$ is the Born-level hard function. These traces give SLL contributions at $\mathcal{O}(\alpha_s^{r_1+r_2+3})$ in perturbation theory. Only insertions of $\Gammac$ give rise to double logarithms, but the remaining structures are necessary to get a non-vanishing color-trace (see~\cite{Becher:2021zkk,Becher:2023mtx} for more details). For this reason the SLLs can arise first at three-loop order. For the $SU(N_c)$ generators in the fundamental representation, i.e.\ (anti-)quark-initiated processes, we can use that
\begin{align} \label{eq_SUN_anticom_fundamental}
    \{\T_i^a, \T_i^b\} &= \frac{1}{N_c} \delta^{ab} \, \Id + \sigma_i \, d^{abc} \, \T_i^c \,; \qquad i=1,2,
\end{align}
with totally symmetric and traceless coefficients $d^{abc}$, to simplify the color traces. The color-space formalism implies that $\sigma_i=+1$ for an initial-state anti-quark and $\sigma_i=-1$ for an initial-state quark. After some algebra one can then show that the traces~\eqref{eq_color_trace_SLL} can be written in terms of a basis consisting of the four operators (in the notation of \cite{Becher:2023mtx})
\begin{align}
    \Oj_2 = (\sigma_1-\sigma_2) \, d^{abc} \, \T_1^a \, \T_2^b \, \T_j^c \,, \qquad
    \Oj_4 = \left(\T_1 - \T_2\right) \cdot \T_j \,, 
\end{align}
and 
\begin{align}
    \S_5 = \T_1 \cdot \T_2 \,, \qquad \S_6 &= \Id \,.
\end{align}
In general, the operators $\Oj_i$ contain a generator for one of the final-state partons (with index $j$), whereas the operators $\S_i$ contain the color generators of the initial-state partons only. The structures $\S_i$ appear when the (real or virtual) soft gluon emitted from $\GammaBar$ is attached to a collinear gluon emitted from $\Gammac$. We remark that more structures appear once gluons in the initial-state (i.e. generators in the adjoint representation) are considered~\cite{Becher:2023mtx}. One can write\footnote{For simplicity, we omit writing ``$\otimes\,\Id$" in the angle brackets from now on.}
\begin{align} \label{eq_color_trace_SLL_evaluated}
    C_{r_1,r_2} = \frac{16}{N_c}\,(-\pi^2) \, (4N_c)^{r_1+r_2+2} \left[ \sum_{j=3}^{2+M} J_j \sum_{i=2,4} c_i^{(r_1)} \tr{\H_{2\to M} \, \Oj_i} + J_{12} \sum_{i=5,6} d_i^{(r_1)} \tr{\H_{2\to M} \, \S_i}\right] ,
\end{align}
with the coefficients $c_i^{(r_1)}$ and $d_i^{(r_1)}$ given by
\begin{align}
\label{eq_CS_coefficients}
    c_2^{(r_1)} &= -\frac{1}{2} \,, & c_4^{(r_1)} &= \frac{2^{-r_1}}{N_c} \,, \nonumber \\
    d_5^{(r_1)} &= -\frac{2}{N_c} \left( 1-2^{-r_1} \right) , & d_6^{(r_1)} &= -\frac{2C_F}{N_c} \, 2^{-r_1} \left( 1-\delta_{0\hspace{0.3mm}r_1} \right) .
\end{align}
The quantities $J_j$ and $J_{12}$ in \eqref{eq_color_trace_SLL_evaluated} are the two relevant angular integrals
\begin{align} \label{eq_angular_integrals}
	J_j &\equiv \int \frac{d\Omega(n_k)}{4\pi} \left(W_{1j}^k-W_{2j}^k\right) \Theta_{\text{veto}}(n_k) \,, \nonumber 
	\\
	J_{12} &\equiv J_2 = -J_1 = \int \frac{d\Omega(n_k)}{4\pi} \, W_{12}^k \, \Theta_{\text{veto}}(n_k) \,,
\end{align}
where $\Theta_{\text{veto}}(n_k) \equiv 1-\Theta_{\text{hard}}(n_k)$ restricts the soft emission to be inside the veto region, and the soft dipole is given by
\begin{align}
    W^k_{ij} = \frac{n_i \cdot n_j}{n_i \cdot n_k \, n_j \cdot n_k} \,.
\end{align}
The set of color traces in~\eqref{eq_color_trace_SLL_evaluated} can easily be calculated for a given hard-scattering process. Evaluating the iterated scale integrals originating from~\eqref{eq_evolution_series} then gives the SLL contribution to the cross section in the double-logarithmic approximation.

\section{Generalized color traces}
\label{sec_generalized_color_trace}

We now turn to the generalized color traces relevant for the resummation of the Glauber series. We define
\begin{align} \label{eq_color_trace_general}
    C_{\{\underline{r}\}}^\ell &\equiv \tr{\H_{2\to M} \left(\Gammac\right)^{r_1} \VG \left(\Gammac\right)^{r_2} \VG \dots \left(\Gammac\right)^{r_{2\ell-1}} \VG \left(\Gammac\right)^{r_{2\ell}} \VG \, \GammaBar } \,,
\end{align}
which for $\ell=1$ reduces to~\eqref{eq_color_trace_SLL}. Recall that each insertion of $\VG$ contributes a factor $i\pi$ to the result. Since the cross section~\eqref{eq_cross_section_general} is a real-valued quantity, there must be an even number of such insertions. The reduction of~\eqref{eq_color_trace_general} to a minimal set of simpler traces is independent of the precise form of the hard function $\H_{2\to M}$. Therefore, we can define
\begin{align}\label{eq:Hexpression}
    \H \equiv \H_{2\to M} \left(\Gammac\right)^{r_1} \VG \left(\Gammac\right)^{r_2} \VG \dots \left(\Gammac\right)^{r_{2\ell-1}} \VG \left(\Gammac\right)^{r_{2\ell}}
\end{align}
as a placeholder to write 
\begin{align}
    C_{\{\underline{r}\}}^\ell &= \tr{\H \, \VG \, \GammaBar} \,,
\end{align}
and then study how the trace reduces under consecutive actions of $\Gammac$ and $\VG$. Application of the first Glauber phase yields
\begin{align} \label{eq_SLL_first_Glauber}
    C_{\{\underline{r}\}}^\ell = 64i\pi \, \sum_{j>2} J_j \, \tr{\H \, \Oj_0} \,,
\end{align}
where the summation index $j$ runs over all final-state partons, including real emissions from $\Gammac$. The new structure
\begin{align} \label{eq_CS_O_0}
    \Oj_0 = if^{abc} \, \T_1^a \, \T_2^b \, \T_j^c
\end{align}
appears only for an odd number of $\VG$ insertions and thus will not appear in the fully reduced form of~\eqref{eq_color_trace_general} presented below.

In deriving a minimal set of simpler traces for an arbitrary number of Glauber phase insertions it is crucial that the set $\{\Oj_0, \Oj_2, \Oj_4, \S_5, \S_6\}$ is closed under repeated applications of $\VG$ and $\Gammac$. This statement will be proven below. Therefore, knowing how $\VG$ and $\Gammac$ act onto these five structures is sufficient to express~\eqref{eq_color_trace_general} in terms of these structures.

\paragraph{Action of $\Gammac$:}
The action of $\Gammac$ differs for color structures with and without $\T_j$. While the virtual part $\Vc_{1,2}\sim\Id$ acts trivially on both types, the real part $\Rc_{1,2}$ can map structures $\Oj_i$ onto $\S_i$ but not vice versa. The action onto an $\S_i$ structure is
\begin{equation} \label{eq_action_Gammac_S}
    \tr{\H \, \Gammac \, \S_i} = 4 \big[ (C_1+C_2) \, \tr{\H \, \S_i} - \tr{\H\,\T_1^a\,\S_i\,\T_1^a} - \tr{\H\,\T_2^a\,\S_i\,\T_2^a} \big] \,,
\end{equation}
where $C_1=C_2=C_F$ for (anti-)quark-initiated processes. This result corresponds to the mappings
\begin{align} \label{eq_GammaC_action_S}
    \S_5 \overset{\Gammac}{\longrightarrow} (4N_c) \, \S_5 \,, \qquad \S_6 \overset{\Gammac}{\longrightarrow} 0 \,.
\end{align}
For the $\Oj_i$ structures the prescription is more complicated and reads
\begin{align}
\label{eq_action_Gammac_Oj}
    \sum_{j>2} J_j \, \tr{\H \, \Gammac \Oj_i} 
    &= 4 \sum_{j>2}{}\strut^{\prime} J_j \Big[ (C_1+C_2)\,\tr{\H\,\Oj_i} - \tr{\H\,\T_1^a\,\Oj_i\,\T_1^a} 
     - \tr{\H\,\T_2^a\,\Oj_i \T_2^a} \Big]
     \nonumber\\
    &\quad + 4\,J_{12}\,\Big[ \tr{\H\,\T_1^{\tilde{a}}\,\Oj_i\,\T_1^a} - \tr{\H\,\T_2^{\tilde{a}}\,\Oj_i\,\T_2^a} \Big] \Big|_{\T_j^b\to-if^{b\tilde{a}a}} \,.
\end{align}
Here the prime on the sum indicates that the gluon whose emission is described by $\Rc_{1,2}$ is excluded. This contribution is taken into account by the second term, in which the replacement $\T_j^b\to-if^{b\tilde{a}a}$ reflects the color structure of the emission of a wide-angle soft gluon from a collinear gluon radiated off parton~1 or 2, see~\cite{Becher:2023mtx} for more details. We find
\begin{align} \label{eq_GammaC_action_O}
    \Oj_0 \overset{\Gammac}{\underset{J_j}{\longrightarrow}} (4N_c) \, \Oj_0 \,, \qquad
    \Oj_2 \overset{\Gammac}{\underset{J_j}{\longrightarrow}} (4N_c) \, \Oj_2 \,, \qquad
    \Oj_4 \overset{\Gammac}{\underset{J_j}{\longrightarrow}} (2N_c) \, \Oj_4 \,,
\end{align}
for the terms in the first row of \eqref{eq_action_Gammac_Oj}, and
\begin{align}\label{eq_GammaC_action_O_part2}
    \Oj_0 \overset{\Gammac}{\underset{J_{12}}{\longrightarrow}} 0 \,, \qquad
    \Oj_2 \overset{\Gammac}{\underset{J_{12}}{\longrightarrow}} 0 \,, \qquad
    \Oj_4 \overset{\Gammac}{\underset{J_{12}}{\longrightarrow}} -(4N_c) (\S_5 + C_F \S_6) \,,
\end{align}
for the terms in the second row. With this result at hand, it is clear that the block $(\Gammac)^{r_{2\ell}}$ in the expression for $\H$ in \eqref{eq:Hexpression} simply results in a factor $(4N_c)^{r_{2\ell}}$ in~\eqref{eq_SLL_first_Glauber}.

\begin{figure}[t]
    \centering
    \includegraphics[scale=1]{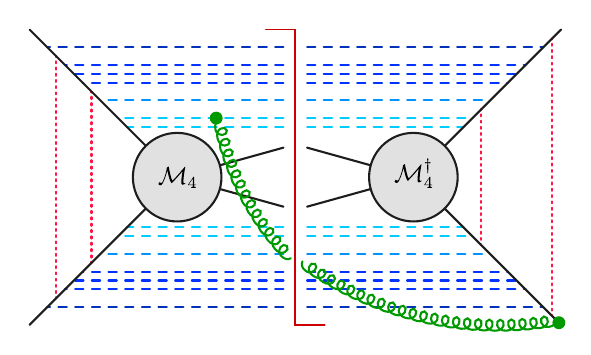}
    \caption{Graphical illustration for a contribution of the $\S_i$ structures to the color traces $C_{\{\underline{r}\}}^{\ell}$ in~\eqref{eq_color_trace_general} relevant for $pp\to 2\,\text{jets}$ production ($M=2$). The soft wide-angle gluon emission is shown in green and can be attached as explained in the main text. Exchanges of Glauber gluons are shown as red dotted lines and collinear gluon emissions as dashed blue lines. Lighter blue colors indicate ``earlier'' collinear emissions (closer to the hard function). This particular diagram shows a contribution to $C_{\{4,2,6,2\}}^2$, which involves $r_1=4$ emissions before the first, $r_2=2$ before the second, $r_3=6$ before the third, and $r_4=2$ emissions before the last Glauber gluon.}
    \label{fig_S_strutures}
\end{figure}

\paragraph{Action of $\VG$:}
The Glauber phase acts onto a generic color structures $\bm{X}$ as
\begin{align}
    \tr{\H \, \VG \bm{X}} = -8i\pi \tr{\,\T_1\cdot\T_2 \, \H \, \bm{X} - \H \, \T_1\cdot\T_2 \, \bm{X}} = -8i\pi \tr{\H \,  \com{\bm{X}}{\T_1\cdot\T_2}\,} \,.
\end{align}
Therefore, one has to calculate the commutator of $\{\Oj_0, \Oj_2, \Oj_4, \S_5, \S_6\}$ with $\T_1\cdot\T_2$. In particular $\VG$ does not mix terms proportional to $J_j$ and $J_{12}$. We find
\begin{align} \label{eq_VG_action_O}
    \Oj_0 &\overset{\VG}{\longrightarrow} -8 i \pi \left(\frac{N_c}{4} \, \Oj_2 - \frac{1}{2}\,\Oj_4\right) , \nonumber
    \\
    \Oj_2 &\overset{\VG}{\longrightarrow} -4i\pi (\sigma_1-\sigma_2)^2 \left(\frac{N_c^2-4}{2N_c}\right) \Oj_0 \,, \nonumber
    \\
    \Oj_4 &\overset{\VG}{\longrightarrow} +16i\pi \, \Oj_0 \,,
\end{align}
where the color coefficient in the second row is related to the cubic Casimir of $SU(N_c)$ in the fundamental representation, $C_{3F}/C_F = \frac{N_c^2 - 4}{2N_c}$. We also obtain
\begin{align} \label{eq_VG_action_S}
    \S_5 \overset{\VG}{\longrightarrow} 0 \,, \qquad
    \S_6 \overset{\VG}{\longrightarrow} 0 \,.
\end{align}
Remarkably, $\VG$ maps the two $\S_i$ structures onto zero. Therefore, the soft gluon emission described by the operator $\GammaBar$ can only be attached to collinear partons emitted from $\Gammac$ insertions before the first Glauber phase (i.e., closest to the hard function), or to one of the original final-state partons, see Figure~\ref{fig_S_strutures} for a graphical illustration. At this point we make two important observations. First, the four color structures $\{\Oj_2, \Oj_4, \S_5, \S_6\}$ that appear already in the color traces~\eqref{eq_color_trace_SLL_evaluated} associated with the SLLs are mapped back onto $\Oj_0$ or 0 under application of $\VG$. This is the key observation to conclude that the generalized color traces~\eqref{eq_color_trace_general} with an arbitrary number of insertions of Glauber operators can be reduced to a finite set of structures. The task of resumming the Glauber phases thus reduces to solving recursive relations. Second, as we already mentioned above, the structure $\Oj_0$ will only show up for an odd number of $\VG$ insertions. This is obvious from~\eqref{eq_GammaC_action_O}, \eqref{eq_GammaC_action_O_part2} and~\eqref{eq_VG_action_O}. Inserting a first Glauber phase creates the color structure $\Oj_0$, while the insertion of an arbitrary number of $\Gammac$ only changes its coefficient. The second Glauber phase, which is always required to get a real-valued cross section, maps $\Oj_0$ onto $\Oj_2$ and $\Oj_4$, so the $\ell=1$ result (the SLL traces) does not depend on $\Oj_0$. Because of the first observation, this argument remains valid for an arbitrary even number of Glauber-operator insertions.

\paragraph{Reduced color traces:}
Applying the above results for the actions of $\Gammac$ and $\VG$, it is possible to reduce the generalized color traces in~\eqref{eq_color_trace_general} for arbitrary $\ell$ and $\{\underline{r}\}$ to
\begin{align}\label{eq_color_trace_general_evaluated}
    C_{\{\underline{r}\}}^{\ell} 
    &= \frac{16}{N_c}\,(-\pi^2)^{\ell}\,(4 N_c)^{n+2\ell} \prod_{i=2}^{\ell} 
     \left[ \frac{(\sigma_1-\sigma_2)^2}{4} \left( 1 - \frac{4}{N_c^2} \right)
     + \frac{2^{2-r_{2i-1}}}{N_c^2} \right] \nonumber\\
    &\quad \times \left[\, \sum_{j=3}^{2+M} J_j \sum_{i=2,4} c_i^{(r_1)} \tr{\H_{2\to M}\,\Oj_i} 
     + J_{12} \sum_{i=5,6} d_i^{(r_1)} \tr{\H_{2\to M}\,\S_i} \right] ,
\end{align}
where $n=\sum_{i=1}^{2\ell} r_i$, and the coefficients $c_i^{(r_1)}$ and $d_i^{(r_1)}$ have been given in~\eqref{eq_CS_coefficients}. Surprisingly, it turns out that the insertion of more Glauber phases just results in an additional prefactor
\begin{align}
   &\,\,\quad (-\pi^2)^{\ell-1}\,(4N_c)^{r_3 + \dots + r_{2\ell}+2(\ell-1)} \prod_{i=2}^{\ell} 
    \left[ \frac{(\sigma_1-\sigma_2)^2}{4}\left(1-\frac{4}{N_c^2}\right)+ \frac{2^{2-r_{2i-1}}}{N_c^2} \right]
    \notag\\
   &= (-\pi^2)^{\ell-1}\,(4N_c)^{r_3 + \dots + r_{2\ell}+2(\ell-1)} 
    \left\{ \begin{array}{cl}
     \displaystyle \left( \frac{4}{N_c^2} \right)^{\ell-1} 2^{-(r_3+r_5+\dots+r_{2\ell-1})} \,;~ 
     & \text{$qq$ or $\bar q\bar q$ scattering,} \\[4mm]
     \displaystyle \prod_{i=2}^{\ell} \left[ 1 - \frac{4}{N_c^2} \left( 1 - 2^{-r_{2i-1}} \right) \right] \,;~
     & \text{$q\bar q$ scattering,}
    \end{array} \right.
\end{align}
compared to the corresponding trace for the case of the SLLs. For $\ell=1$ this factor reduces to 1 and we recover the result~\eqref{eq_color_trace_SLL_evaluated}. We observe that only for $\sigma_1=\sigma_2$, i.e.\ for $qq$ or $\bar q\bar q$ scattering, higher Glauber phases are suppressed in the large-$N_c$ limit. However, this is not the case for $q\bar q$ scattering, for which $\sigma_1=-\sigma_2$. 

\paragraph{Comment on the color basis:}
It is interesting to think about possible color structures that can in general appear. On the one hand, $\Gammac$ and $\VG$ only depend on $\T_1$ and $\T_2$, so it is obvious from~\eqref{eq_SLL_first_Glauber} that all color structures can only contain $\Id_j$ or $\T_j$ for $j>2$. On the other hand, for generators in the (anti-)fundamental representation of $SU(N_c)$, i.e. for (anti-)quarks in the initial-state, it is always possible to reduce to product of two generators \begin{align}
    \T_i^a \, \T_i^b = \frac{1}{2} \left(\com{\T_i^a}{\T_i^b} + \anticom{\T_i^a}{\T_i^b} \right) = \frac{\delta^{ab}}{2N_c} \, \Id_i + \frac{1}{2}\left(if^{abc} + \sigma_i \, d^{abc}\right) \T_i^c \,; \qquad i=1,2 \,.
\end{align}
Therefore, all color structures can only contain $\Id_i$ or $\T_i$ for $i=1,2$. Those two constrains allow for the following seven structures
\begin{align}
	if^{abc} \, \T_1^a \, \T_2^b \, \T_j^c \,, \qquad \sigma_1 \,d^{abc} \, \T_1^a \, \T_2^b \, \T_j^c \,, \qquad 
	 \sigma_2 \,d^{abc} \, \T_1^a \, \T_2^b \, \T_j^c \,, \nonumber\\ 
	\Id_1 \, \T_2^a\, \T_j^a \,, \qquad \T_1^a \, \Id_2 \, \T_j^a \,, \qquad \T_1^a \, \T_2^a \, \Id_j \,, \qquad
	 \Id_1 \, \Id_2 \, \Id_j \,. \hspace{5mm} 
\end{align}
Since the cross section has to be invariant under the relabeling $1\leftrightarrow2$ and the angular integrals in~\eqref{eq_angular_integrals} transform as
\begin{align}
    J_j \to - J_j \,, \qquad J_{12} \to + J_{12}
\end{align}
under this exchange, the structures $\Oj_i$ need to be anti-symmetric and the structures $\S_i$ need to be symmetric. Therefore, we are left with the structures $\{\Oj_0, \Oj_2, \Oj_4, \S_5, \S_6\}$ as a basis for the problem at hand. This simple argument proves that one can simplify the generalized color traces~\eqref{eq_color_trace_general} to a linear combination of those five basis structures. It is nevertheless an interesting observation and quite surprising that the result for such a complicated problem is actually as simple as~\eqref{eq_color_trace_general_evaluated}.

\section{Nested scale integrals and partonic cross section}
\label{sec_nested_integrals}

In addition to the color-traces $C_{\{\underline{r}\}}^{\ell}$ the coefficients of the Glauber series are determined by nested scale integrals from the expansion of the path-ordered exponential in~\eqref{eq_evolution_series}. To make the structure of these integrals more transparent, we define
\begin{align}\label{eq:scale_ints}
    I_r[f](\mu_h, \mu) &\equiv \int_{\mu}^{\mu_h} \frac{d\mu_0}{\mu_0} \, \frac{\alpha_s(\mu_0)}{4\pi} \, \int_{\mu_0}^{\mu_h} \frac{d\mu_1}{\mu_1} \, \frac{\alpha_s(\mu_1)}{4\pi} \, \ln \frac{\mu_1^2}{\mu_h^2} \, \dots \int_{\mu_{r-1}}^{\mu_h} \frac{d\mu_r}{\mu_r} \, \frac{\alpha_s(\mu_r)}{4\pi} \, \ln \frac{\mu_r^2}{\mu_h^2} \, f(\mu_h,\mu_r) \,,
\end{align}
where $f$ is an arbitrary smooth function. The integrals in $I_r$ result from inserting the block $(\Gammac)^r \, \VG$ in the color trace, and thus only the integrations over $\mu_1,\dots,\mu_r$ contain an additional logarithm from the cusp anomalous dimension. We can then express the full nested scale integral associated with the color trace~\eqref{eq_color_trace_general} as the composition
\begin{align}
\label{eq_nested_integrals}
    I_{\{\underline{r}\}}^{\ell}(\mu_h,\mu_s)=(I_0 \circ I_{r_{2\ell}} \circ I_{r_{2\ell-1}} \circ \dots \circ I_{r_2} \circ I_{r_1})[1](\mu_h, \mu_s) \,,
\end{align}
where the last integration $I_0$ originates from $\GammaBar$. Note that the ordering of the integrals is opposite to that of the color operators in \eqref{eq:Hexpression}, as follows from \eqref{eq_evolution_series}.

In the strict double-logarithmic approximation (with the counting $\pi = |\ln(-1)| \sim L$) the running of the strong coupling $\alpha_s(\mu)$ can be neglected. The integrals~\eqref{eq_nested_integrals} can then be expressed in a closed form by using that
\begin{align}
    I_r[\ln^a](\mu_h,\mu) = \left(\frac{\alpha_s(\bar{\mu})}{4\pi}\right)^{r+1}
     \frac{(-2)^r}{1+2r+a}\,\frac{a!!}{(2r+a)!!}\,\ln^{1+2r+a}\!\Big(\frac{\mu_h}{\mu}\Big)
\end{align}
when $I_r$ acts on a pure power of the logarithm $\ln(\mu_h/\mu)$. Here $n!! = n (n-2) (n-4) \dots$ denotes the double factorial, and $\bar{\mu}$ is an arbitrary fixed reference scale between $\mu_s$ and $\mu_h$. We then obtain 
\begin{align}
\label{eq_nested_integrals_evaluated}
    I_{\{\underline{r}\}}^{\ell}(\mu_h,\mu_s) 
    = \left(\frac{\alpha_s(\bar{\mu})}{4\pi}\right)^{2\ell+n+1} 
     \frac{(-2)^n\,L^{2n+2\ell+1}}{(2n+2\ell)(2n+2\ell+1)}\,\prod_{k=1}^{2\ell}\,
     \frac{(2\sum_{i=1}^{k-1}r_i+k-3)!!}{(2\sum_{i=1}^{k}r_i+k-1)!!} \,,
\end{align}
where now $L=\ln(\mu_h/\mu_s)$, and we define $(-2)!!\equiv (-1)!!\equiv 1$. 

We now have everything at hand to present our main result. In the leading double-logarithmic approximation, and including an arbitrary number of Glauber phases, the contribution of the Glauber series (consisting of SLLs plus all higher-order Glauber terms) to the cross section for a $2\to M$ jet process can be written as
\begin{equation}
   \sigma_{2\to M}^{\rm SLL+G}(Q_0) 
   = \sum_{i \in \{q,\bar q,g\}} \int\!dx_1 \int\!dx_2\,f_1(x_1,\mu_s)\,f_2(x_2,\mu_s)\,\sum_{\ell=1}^{\infty} \sum_{\{\underline{r}\}} I_{\{\underline{r}\}}^{\ell}(\mu_h,\mu_s)\,C_{\{\underline{r}\}}^{\ell} \,.
\end{equation}
Here the sum runs over {\it all} partonic channels $1+2\to 3+\dots+(2+M)$, and it is implicitly understood that the $C_{\{\underline{r}\}}^{\ell}$ depend on a given partonic reaction. Clearly, the restriction to quarks or anti-quarks in the initial state would not be justified if we would aim at precision phenomenology for such an observable. (The corresponding analysis for gluons in the initial-state is considerably more complicated and will be presented elsewhere.) Therefore, in this work we are interested in estimating the size of the Glauber phases for a given {\em partonic\/} scattering process $\hat\sigma_{2\to M}$, for which we obtain
\begin{align}
\label{eq_rlsums}
   \hat{\sigma}_{2\to M}^{\rm SLL+G}(Q_0) 
   = \sum_{\ell=1}^{\infty} \sum_{\{\underline{r}\}} I_{\{\underline{r}\}}^{\ell}(\mu_h,\mu_s)\,
    C_{\{\underline{r}\}}^{\ell}
   \equiv \sum_{\ell=1}^{\infty} \sum_{r_1=0}^{\infty} \sum_{r_2=0}^{\infty} \dots \sum_{r_{2\ell}=0}^{\infty} 
    \hat{\sigma}_{\{\underline{r}\}}^{\ell} \,.
\end{align}
In Section~\ref{sec_pheno} we will study the numerical size of this contribution relative to the Born cross section for a few selected processes. Combining~\eqref{eq_CS_coefficients},~\eqref{eq_color_trace_general_evaluated} and~\eqref{eq_nested_integrals_evaluated} we now obtain the master formula for the all-order coefficients of the Glauber series for quark-initiated processes
\begin{align} \label{eq_master_formula}
    \hat{\sigma}_{\{\underline{r}\}}^{\ell} 
    &= \frac{4\alpha_s(\bar{\mu}) L}{\pi N_c} \frac{2^{n} \left(-w\right)^{\ell+n} w_\pi^{\ell} }{(2n+2\ell)(2n+2\ell+1)} \, \prod_{k=1}^{2\ell} \frac{(2\sum_{i=1}^{k-1}r_i+k-3)!!}{(2\sum_{i=1}^{k}r_i+k-1)!!} \nonumber \\[0.4ex]
    &\quad \times \prod_{i=2}^{\ell} \left[ \frac{(\sigma_1-\sigma_2)^2}{4}\left(1-\frac{4}{N_c^2}\right)+ \frac{2^{2-r_{2i-1}}}{N_c^2}\right] \nonumber
    \\
    &\quad \times \Bigg[\, \sum_{j=3}^{2+M} J_j \left(-\frac{1}{2}\,\tr{\H_{2\to M} \, \Oj_2} + \frac{2^{-r_1}}{N_c} \tr{\H_{2\to M} \, \Oj_4}\right) \nonumber \\[-2mm]
    &\hspace{14mm} - \frac{2}{N_c} \, J_{12} (1-\delta_{0r_1}) \Big((1-2^{-r_1}) \tr{\H_{2\to M} \, \S_5} + C_F \, 2^{-r_1} \tr{\H_{2\to M} \, \S_6} \Big) \Bigg] \,,
\end{align}
with $n=\sum_{i=1}^{2\ell} r_i$ and $\mathcal{O}(1)$ expansion parameters as defined in~\eqref{eq_wwpi}. Again, for $\ell=1$ we recover the corresponding SLL result. In this simpler case, the sums in~\eqref{eq_rlsums} reduce to a double sum, which can be evaluated analytically \cite{Becher:2023mtx}. What makes the problem at hand considerably more complicated is not necessarily the lenghty structure in~\eqref{eq_master_formula}, but rather the infinitely many infinite sums in~\eqref{eq_rlsums}. Nevertheless, for some special cases closed analytic expressions can be derived. 

One example is the case $n=0$ for arbitrary $\ell$, which corresponds to summing insertions of $\VG$ \emph{without\/} summing the double-logarithms from $\Gammac$. In that case we have $r_i=0$ for all $i$, and the only remaining sum over $\ell$ can be performed in terms of simple trigonometric functions. We obtain
\begin{align}
\label{eq_resummedVG_0}
    \sum_{\ell=1}^{\infty} \hat{\sigma}_{\{\underline{0}\}}^{\ell} 
    = \frac{4\alpha_s(\bar{\mu}) L}{\pi N_c} \, \frac{1}{\kappa_0} 
     \! \left(\frac{\sin \sqrt{\kappa_0 ww_{\pi}}}{\sqrt{\kappa_0 ww_{\pi}}}-1\right) \!
     \sum_{j=3}^{2+M} J_j \left(-\frac{1}{2}\,\tr{\H_{2\to M} \, \Oj_2} + \frac{1}{N_c}\,\tr{\H_{2\to M} \, \Oj_4}\right) ,
\end{align}
with the coefficient
\begin{align}
    \kappa_0 = \left\{ \begin{array}{cl}
     \displaystyle \frac{4}{N_c^2} \,;~ & \text{$qq$ or $\bar q\bar q$ scattering,} \\[4mm]
     1 \,;~ & \text{$q\bar q$ scattering,}
    \end{array} \right.
\end{align}
which is a special case of
\begin{align}
\label{eq_kappar}
    \kappa_r = \frac{(\sigma_1-\sigma_2)^2}{4}\left(1-\frac{4}{N_c^2}\right)+ \frac{2^{2-r}}{N_c^2} \,.
\end{align}
It turns out that the two traces in~\eqref{eq_resummedVG_0} vanish at tree-level for $2\to 1$ and $2\to 0$ partonic scattering processes. This motivates us to consider also the terms in the Glauber series containing a single insertion of $\Gammac$, corresponding to $n=1$. We obtain
\begin{align}
\label{eq_resummedVG_1}
    \sum_{\ell=1}^{\infty} \, \sum_{\{r_i\},\, n = 1} \hat{\sigma}_{\{\underline{r}\}}^{\ell} 
    &= \frac{4\alpha_s(\bar{\mu}) L}{\pi N_c} \, \frac{2w}{\kappa_0} \left(\frac{\sin \sqrt{\kappa_0ww_{\pi}}}{(\kappa_0ww_{\pi})^{3/2}}-\frac{1}{\kappa_0ww_{\pi}}+\frac{1}{6}\right) \nonumber \\
    &\quad \times \Bigg[\, \sum_{j=3}^{2+M} J_j \left(-\frac{1}{2}\,\tr{\H_{2\to M} \, \Oj_2} + \frac{1}{2N_c}\,\tr{\H_{2\to M} \, \Oj_4}\right) \nonumber \\[-2mm]
    &\hspace{14mm} - \frac{1}{N_c}\,J_{12} \Big(\tr{\H_{2\to M} \, \S_5} + C_F \, \tr{\H_{2\to M} \, \S_6} \Big) \Bigg] \nonumber \\
    &- \frac{4\alpha_s(\bar{\mu}) L}{\pi N_c} \, \frac{w}{2\kappa_0} \Bigg[ (\kappa_0ww_{\pi}-3)\,\frac{\sin \sqrt{\kappa_0ww_{\pi}}}{(\kappa_0ww_{\pi})^{3/2}} + 3\,\frac{\cos \sqrt{\kappa_0ww_{\pi}}}{\kappa_0ww_{\pi}} \nonumber \\
    &\quad+ \frac{\kappa_1}{\kappa_0} \left((\kappa_0ww_{\pi}-5)\,\frac{\sin \sqrt{\kappa_0ww_{\pi}}}{(\kappa_0ww_{\pi})^{3/2}}+5\,\frac{\cos \sqrt{\kappa_0ww_{\pi}}}{\kappa_0ww_{\pi}}+\frac{2}{3}\right) \Bigg]
    \nonumber \\
    &\quad \times \sum_{j=3}^{2+M} J_j \left(-\frac{1}{2}\,\tr{\H_{2\to M} \, \Oj_2} + \frac{1}{N_c} \, \tr{\H_{2\to M} \, \Oj_4}\right) .
\end{align}
The sum of the expressions~\eqref{eq_resummedVG_0} and~\eqref{eq_resummedVG_1} is still not a particularly good numerical approximation, because the double-logarithmic corrections arising in yet higher orders are typically important \cite{Becher:2023mtx}. However, once we include the $n=2$ term we do obtain a closed analytic formula which is numerically very close to the full resummed result for all the partonic channels considered below, see Figure~\ref{fig_qq_qq_l_analytic} for an example. The respective analytic formula is given in Appendix~\ref{app_Glauber_phases_resummed_n_2}. 

We lastly mention that the ambiguity in the choice of the reference scale $\bar{\mu}$ can be avoided by including the running of the QCD coupling $\alpha_s(\mu)$ under the scale integrals in \eqref{eq:scale_ints}. For example,~\eqref{eq_resummedVG_0} can then be trivially generalized by making the replacement 
\begin{align}
    \alpha_s(\bar{\mu})L \to \frac{2\pi}{\beta_0} \ln\frac{\alpha_s(Q_0)}{\alpha_s(Q)} \,,
\end{align}
where $\beta_0$ is the one-loop coefficient of the QCD $\beta$-function. The generalization of~\eqref{eq_resummedVG_1} and~\eqref{eq_resummedVG_2}, however, is more complicated. As discussed in~\cite{Becher:2023mtx}, using one-loop running of $\alpha_s(\mu)$ can be approximated well by choosing the fixed reference scale as the geometric mean, $\bar{\mu} = \sqrt{QQ_0}$. Therefore, we use formula~\eqref{eq_master_formula} with fixed coupling for the numerical predictions presented below.

\section{Phenomenological implications}
\label{sec_pheno}

To analyze the Glauber series we need to evaluate the color traces appearing in~\eqref{eq_master_formula}. In this work we only consider processes with one color structure contributing to the amplitude. As a consequence, our analysis simply results in an overall correction factor that multiplies the Born cross section. We remark, however, that in general the terms in the Glauber series are shape dependent. We express the leading-order hard function of a given partonic process generically as
\begin{align}
\label{eq_H_tensor}
    \H_{2\to M} &= \langle\H_{2\to M}\rangle \; \mathcal{T}_{\alpha_1\dots\alpha_{2+M}} \, \mathcal{T}^{\dagger}_{\beta_1\dots\beta_{2+M}} \,,
\end{align}
where ${\cal T}_{\alpha_1 \dots \alpha_{2+M}}$ denotes the color tensor associated with the amplitude $|\mathcal{M}_{2+M}\rangle$, and ${\cal T}^\dagger_{\beta_1\dots\beta_{2+M}}$ that associated with the complex conjugate amplitude $\langle\mathcal{M}_{2+M}|$, respectively. The indices $\alpha_i$ and $\beta_i$ are fundamental indices of $SU(N_c)$ for parton $i$ an (anti-)quark, or adjoint indices for a gluon. We remark that the trace on the right-hand side in~\eqref{eq_H_tensor} does \emph{not} contain the angular integrations denoted by~$\otimes$,
but they are required to normalize the hard functions to the Born cross section
\begin{align} \label{eq_normalization_hard_function}
    \tr{\H_{2\to M} \otimes \Id} &= \hat{\sigma}_{2\to M} \,.
\end{align}
This then obviously determines the color traces involving the basis structure \mbox{$\S_6 = \Id$} in all cases.

In the following we estimate the size of Glauber phases for different quark-initiated $2\to2$, $2\to1$ and $2\to0$ processes. As~\eqref{eq_master_formula} is expressed in terms of infinitely many infinite sums, which in general cannot be performed in a closed analytic form, we instead study the convergence of the Glauber series by defining the following partial sums:
\begin{itemize}
\item[(A)] 
For fixed power $\ell = 1,2, \dots$ of the variable $w_\pi$ in~\eqref{eq_master_formula}, we perform the sum over $n$ in the logarithmically-enhanced contributions proportional to $w^{\ell+n}$. The sum over $\ell$ is taken in a second step. The Glauber series is then organized in the form
\begin{align} \label{eq_Glauberseries_A}
   \frac{\hat{\sigma}_{2\to M}^{\rm SLL+G}}{\hat{\sigma}_{2\to M}} 
   \sim w_{\pi} \left( -c_{1,0}\,w + c_{1,1}\,w^2 \mp \dots \right) 
    + w_{\pi}^2 \left( c_{2,0}\,w^2 - c_{2,1}\,w^3 \pm \dots \right) + \dots
\end{align}
with positive coefficients $c_{\ell,n}$.  
\item[(B)] 
For fixed power $N=\ell+n=1,2,\dots$ of the variable $w$ in~\eqref{eq_master_formula}, we perform the sum over $\ell=1,\dots,N$, thus summing the series of terms involving powers of the variable $w_\pi$. The sum over $N$ is taken in a second step. The Glauber series is then organized in the form
\begin{align} \label{eq_Glauberseries_B}
   \frac{\hat{\sigma}_{2\to M}^{\rm SLL+G}}{\hat{\sigma}_{2\to M}}
   \sim -c_{1,0}\,ww_{\pi} + w^2 \left(c_{1,1} \, w_{\pi}+c_{2,0}\,w_{\pi}^2\right) 
    - w^3 \left(c_{1,2}\,w_{\pi}+c_{2,1}\,w_{\pi}^2+c_{3,0}\,w_{\pi}^3\right) \pm \dots \,.
\end{align}
Note that this expansion does \emph{not} coincide with the expressions~\eqref{eq_resummedVG_0} and~\eqref{eq_resummedVG_1}, because each insertion of a Glauber phase yields a factor of $\sqrt{w w_\pi}$.
\item[(C)] 
Lastly, we treat $w$ and $w_\pi$ on the same footing and fix $\mathcal{N}=n+2\ell=2,3,\dots$ in~\eqref{eq_master_formula} while performing the sum over $n=0,\dots,(\mathcal{N}-2)$. The sum over $\mathcal{N}$ is taken in a second step. The Glauber series is then organized in the form
\begin{align} \label{eq_Glauberseries_C}
   \frac{\hat{\sigma}_{2\to M}^{\rm SLL+G}}{\hat{\sigma}_{2\to M}} 
   \sim -c_{1,0}\,ww_{\pi} + c_{1,1}\,w^2w_{\pi} - \left(c_{1,2}\,w^3w_{\pi} - c_{2,0}\,w^2w_{\pi}^2\right) \pm \dots \,.
\end{align}
\end{itemize}
We will also compare the expressions~\eqref{eq_resummedVG_0},~\eqref{eq_resummedVG_1} and~\eqref{eq_resummedVG_2} against the ``all-order'' (in practice this means a truncation of the multi-sum for $\ell = 3$ and $n = 20$) result in~\eqref{eq_master_formula}. In all plots below we use the two-loop running coupling with $\alpha_s(M_Z)=0.118$, and following~\cite{Becher:2023mtx} we include contributions of the two-loop cusp anomalous dimension by replacing
\begin{align}
    \alpha_s(\bar{\mu}) \to \left(1+\frac{\gamma_1^{\text{cusp}}}{\gamma_0^{\text{cusp}}}\,\frac{\alpha_s(\bar{\mu})}{4\pi}\right)\alpha_s(\bar{\mu}) \,.
\end{align}

\subsection{Numerical estimates for \texorpdfstring{$\bm{2\to 2}$}{2->2} processes}
\label{subsec_2to2}

\begin{figure}[t!]
    \centering
    \includegraphics[scale=1]{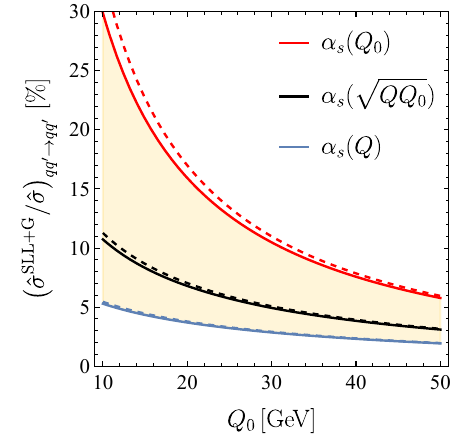}
    \includegraphics[scale=1]{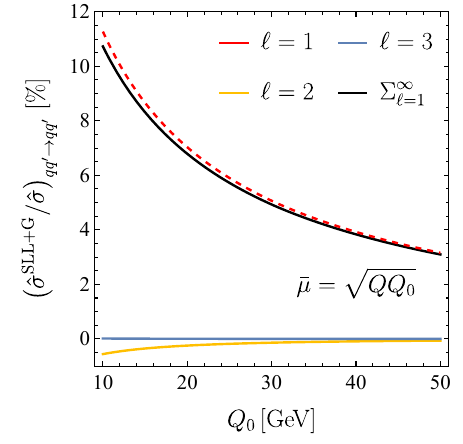}
    \\
    \includegraphics[scale=1]{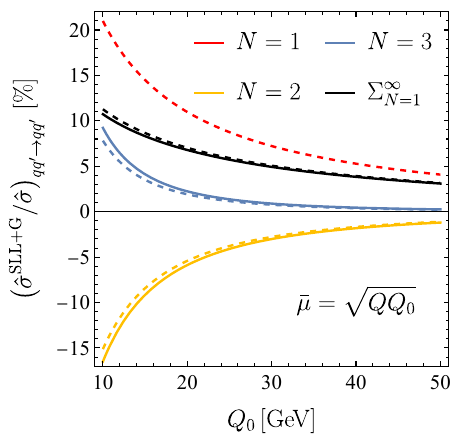}
    \includegraphics[scale=1]{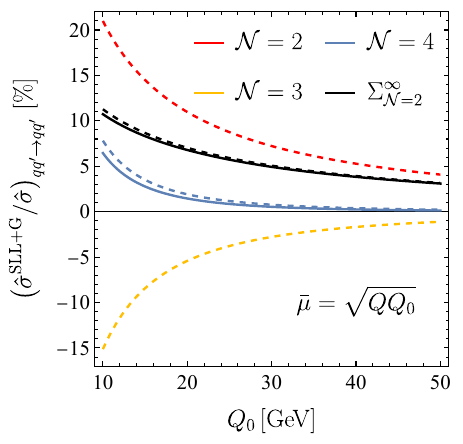}
    \caption{Numerical estimates of the impact of the Glauber series on $qq'\to qq'$ scattering. The upper left figure shows the all-order result obtained with three different choices for the scale in the running coupling. The remaining three figures show the partial sums defined by the scenarios A (upper right panel), B (bottom left panel) and C (bottom right panel). Dashed lines, whenever plotted, show the corresponding curves in the approximation of keeping only the SLLs ($\ell=1$). In cases where solid and dashed lines agree, we only show the dashed curves. In all plots we use $Q=1$\,TeV and $\Delta Y=2$, as well as $\bar{\mu}=\sqrt{QQ_0}$ in all but the first plot. Note that the all-order result (black curve) is the same in all four cases.}
    \label{fig_qq_qq}
\end{figure}

We consider forward scattering of two quarks of different flavor $qq'\to qq'$ via gluon exchange, such that $\sigma_1-\sigma_2=0$. The leading-order hard function then reads
\begin{align}
    \H_{qq'\to qq'} = \langle\H_{qq'\to qq'}\rangle \, \frac{4}{N_c^2-1} \, (t^a)_{\alpha_3\alpha_1} (t^a)_{\alpha_4\alpha_2} (t^b)_{\beta_1\beta_3} (t^b)_{\beta_2\beta_4} \,,
\end{align}
which, in addition to the normalization condition~\eqref{eq_normalization_hard_function}, results in the non-vanishing color traces
\begin{align}
	\sum_{j=3}^4 J_j\, \tr{\H_{qq'\to qq'} \, \Oj_4} 
	&= - C_F \, \hat{\sigma}_{qq'\to qq'} \, J_{43} \,, \notag\\[-2mm]
	\tr{\H_{qq'\to qq'} \, \S_5} &= - \frac{1}{N_c} \, \hat{\sigma}_{qq'\to qq'} \,,
\end{align}
where $J_{43}\equiv J_4-J_3$. Figure~\ref{fig_qq_qq} shows our numerical estimates of the effects from the Glauber phases for $qq'\to qq'$ forward scattering with $J_{43} = 2 J_{12} = 2 \Delta Y$. We observe that the net effect of the Glauber series turns out to be very small compared to the resummed SLLs. This is somewhat surprising, given the relatively large numerical value of the expansion parameter $w_\pi$, see the discussion below~\eqref{eq_wwpi}. The smallness of the net result is mainly due to two different reasons. First, the numerical coefficients in~\eqref{eq_master_formula} turn out to be small. And second, the factor $(-1)^{\ell+n}$ leads to a cancellation of potentially large terms in the alternating sum. 

To highlight the last point we compare the convergence of the Glauber series adopting the different partial sums defined in scenarios A\,--\,C explained above. In scenario A (upper right panel in Figure~\ref{fig_qq_qq}) this cancellation happens for each term in the sum over $\ell$, see~\eqref{eq_Glauberseries_A}. Hence, the values obtained for $\ell=2$ are more than an order of magnitude small than those for $\ell=1$. The situation is different in scenarios B and C (lower panels of Figure~\ref{fig_qq_qq}), which turn out to be numerically similar to each other due to the smallness of the coefficients $c_{\ell,n}$ for $\ell \geq 2$. In these schemes higher-order terms in the Glauber series play an important role, but the sum of all terms is the same as before. We hence conclude that summing all logarithmically-enhanced corrections for fixed powers of $w_\pi$ (scenario A) leads to a rapidly converging Glauber series, and that in this case it is numerically sufficient to include two Glauber phases, which is the SLL series. The same conclusion holds for all other partonic channels considered in this work.

\begin{figure}[t]
    \centering
    \includegraphics[scale=1]{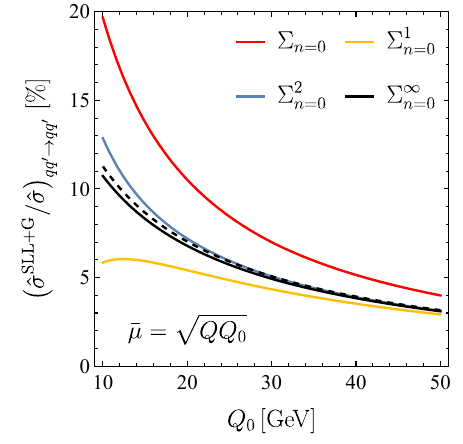}
    \caption{Comparison of the closed analytic expressions~\eqref{eq_resummedVG_0},~\eqref{eq_resummedVG_1} and~\eqref{eq_resummedVG_2} to the cross section for $qq'\to qq'$ scattering. Again, the dashed line shows the SLL contribution obtained with only two Glauber-gluon exchanges, and we use $Q = 1\,\text{TeV}$ and $\Delta Y = 2$ as well as $\bar{\mu}=\sqrt{QQ_0}$.}
    \label{fig_qq_qq_l_analytic}
\end{figure}

In Figure~\ref{fig_qq_qq_l_analytic} we compare the analytic approximations~\eqref{eq_resummedVG_0},~\eqref{eq_resummedVG_1} and~\eqref{eq_resummedVG_2} with the all-order result. Recall that these approximations amount to the summation of all Glauber phases $\VG$ for a fixed number of $\Gammac$ insertions. As the insertion of a Glauber phase is $\sim\!\sqrt{w w_\pi}$ this does \emph{not} correspond to one of the three scenarios A\,--\,C. Figure~\ref{fig_qq_qq_l_analytic} shows that the analytic expression with the sum over $n = 0,1,2$ shows good agreement with the all-order result over a wide range of physically relevant values of $Q_0$. Eventually, however, these analytic expressions diverge from the black curve for small enough values of $Q_0$.

\begin{figure}[t]
    \centering
    \includegraphics[scale=1]{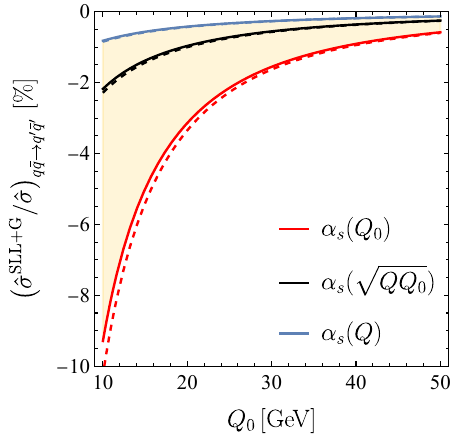}
    \includegraphics[scale=1]{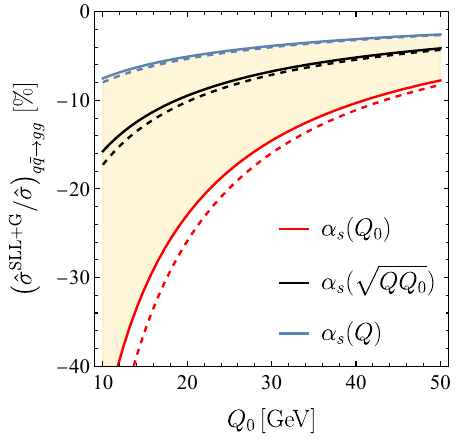}
    \\
    \includegraphics[scale=1]{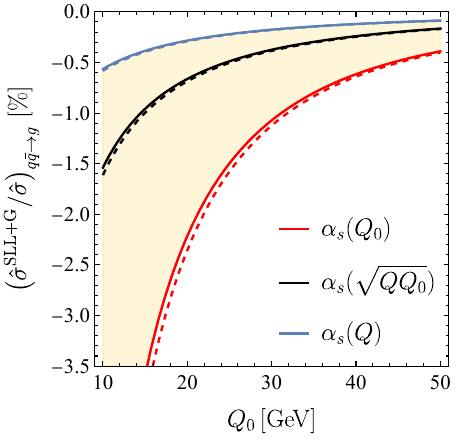}
    \includegraphics[scale=1]{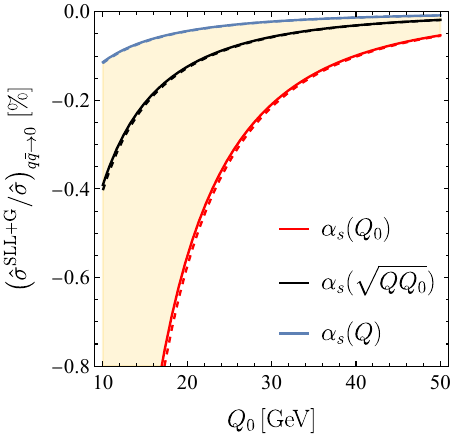}
    \caption{Numerical estimates of the impact of the Glauber series on $q\bar{q}\to q'\bar{q}'$ (upper left panel), $q\bar{q}\to gg$ (upper right panel), $q\bar{q}\to g$ (bottom left panel) and $q\bar{q}\to 0$ (bottom right panel) scattering for different choices of the reference scale $\bar{\mu}$. Dashed lines show the corresponding curves in the SLL limit. We use $Q = 1\,\text{TeV}$ and $\Delta Y = 2$.}
    \label{fig_qqbar}
\end{figure}

We further consider the $q\bar q$ annihilation into a quark anti-quark pair of a different flavor, $q\bar{q}\to q'\bar{q}'$, and the annihilation into two gluons, $q\bar{q}\to gg$, for which $\sigma_1=-1$ and $\sigma_2=1$.
For the former case the leading-order hard function reads
\begin{align}
    \H_{q\bar{q}\to q'\bar{q}'} = \langle\H_{q\bar{q}\to q'\bar{q}'}\rangle \, \frac{4}{N_c^2-1} \, (t^a)_{\alpha_2\alpha_1} (t^a)_{\alpha_3\alpha_4} (t^b)_{\beta_1\beta_2} (t^b)_{\beta_4\beta_3} \,,
\end{align}
and the non-trivial color traces evaluate to
\begin{align}
	\sum_{j=3}^4 J_j \, \tr{\H_{q\bar{q}\to q'\bar{q}'} \, \Oj_2} 
	&= \frac{N_c^2-4}{N_c^2}\,\hat{\sigma}_{q\bar{q}\to q'\bar{q}'} \, J_{43} \,,  \nonumber\\
	\sum_{j=3}^4 J_j \tr{\H_{q\bar{q}\to q'\bar{q}'} \, \Oj_4} 
	&= \frac{N_c^2-4}{2N_c}\,\hat{\sigma}_{q\bar{q}\to q'\bar{q}'} \, J_{43} \,,  \nonumber\\
	\tr{\H_{q\bar{q}\to q'\bar{q}'} \, \S_5} &= \frac{1}{2N_c} \,\hat{\sigma}_{q\bar{q}\to q'\bar{q}'} \,.
\end{align}
In the foward-scattering limit, the process $q\bar{q}\to gg$ is dominated by the $t$-channel quark-exchange diagram. The leading-order hard function is then given by 
\begin{align}
    \H_{q\bar{q}\to gg} = \langle\H_{q\bar{q}\to gg}\rangle \, \frac{1}{C_F^2 N_c} \, (t^{a_4}t^{a_3})_{\alpha_2\alpha_1} (t^{b_3}t^{b_4})_{\beta_1\beta_2} \,,
\end{align}
and the non-vanishing color traces are
\begin{align}
	\sum_{j=3}^4 J_j\,\tr{\H_{q\bar{q}\to gg} \, \Oj_4} 
	&= \frac{N_c^2}{4C_F} \, \hat{\sigma}_{q\bar{q}\to gg} \, J_{43} \,,  \notag\\[-2mm]
	\tr{\H_{q\bar{q}\to gg} \, \S_5} &= -\frac{1}{4N_c^2 C_F} \,\hat{\sigma}_{q\bar{q}\to gg} \,.
\end{align}
For both processes similar conclusions hold. The net effects of the all-order Glauber series are shown in the upper row of Figure~\ref{fig_qqbar}. We note that for $q\bar{q}\to q'\bar{q}'$ the absolute effect of the Glauber series is significantly smaller due to a cancellation of contributions from $\Oj_2$ and $\Oj_4$ for $n=0$. This effectively means that the series starts at four-loop order in our approximation. In contrast, for the gluonic channel $q\bar{q} \to gg$ the contributions are somewhat larger due to larger color factors for gluons. Nevertheless, also in this case the net effect of the Glauber series is negligible compared to the SLLs. 

\subsection{Numerical estimates for \texorpdfstring{$\bm{2\to 1}$}{2->1} and \texorpdfstring{$\bm{2\to 0}$}{2->0} processes}
\label{subsec_2to1and0}

The tree-level hard function for the partonic process $q\bar{q} \to g$ reads
\begin{align}
    \H_{q\bar{q}\to g} = \langle\H_{q\bar{q}\to g}\rangle \, \frac{2}{N_c^2-1} \,  (t^{a_3})_{\alpha_2\alpha_1} (t^{b_3})_{\beta_1\beta_2} \,.
\end{align}
Again $\sigma_1=-1$ and $\sigma_2=1$, and the only non-vanishing trace besides the normalization~\eqref{eq_normalization_hard_function} is
\begin{align}
	\tr{\H_{q\bar{q}\to g} \, \S_5} &= \frac{1}{2N_c} \,\hat{\sigma}_{q\bar{q}\to g} \,.
\end{align}
For quark anti-quark scattering into colorless final states, such as $q\bar{q} \to \gamma, Z^0, W^\pm$, we have $\sigma_1=-1$ and $\sigma_2=1$ and the hard function reads 
\begin{align}
    \H_{q\bar{q}\to 0} = \langle\H_{q\bar{q}\to 0}\rangle \, \frac{1}{N_c} \, \delta_{\alpha_2\alpha_1} \delta_{\beta_1\beta_2} \,.
\end{align}
As there are no colored particles in the final state, the sum over $j\geq3$ in the second-to-last line in~\eqref{eq_master_formula} is absent. The only relevant color trace besides the normalization is
\begin{align}
	\tr{\H_{q\bar{q}\to 0} \, \S_5} &= -C_F \, \hat{\sigma}_{q\bar{q}\to 0} \,.
\end{align}
When considering $2\to 1$ and $2\to 0$ processes at least one color-neutral particle needs to participate in the scattering. Such particles are always included in the hard functions $\H_m$ and in the phase-space integrations represented by the $\otimes$ symbol. Numerical estimates of the net effect of the Glauber series are shown in the bottom left panel of Figure~\ref{fig_qqbar} for $q\bar{q} \to g$ scattering and in the bottom right panel for $q\bar{q} \to 0$ scattering. Again, similar conclusions as for $qq' \to qq'$ scattering hold for these partonic channels as well. Note that the overall impact of SLLs and the Glauber series on $2\to 1$ and $2\to 0$ processes are strongly suppressed compared to $2\to 2$ scattering processes. The reason is that the leading terms start at four- and five-loop order in these cases, respectively, rather than at three loops. This implies a larger suppression from the numerical coefficients in~\eqref{eq_master_formula}.

\section{Conclusions}
\label{sec_conclusions}

The perturbative series for non-global observables at hadron colliders receive enhanced double-logarithmic corrections in higher orders. These super-leading logarithms (SLLs) are generated by complex phases in the amplitude, arising from soft Glauber-gluon exchanges between the colliding partons. Higher-order Glauber exchanges are associated with higher powers of $\pi^2$ and \textit{a priori} are numerically not suppressed compared with the SLLs for experimentally relevant values of $Q$ and $Q_0$. Both contributions give rise to $\mathcal{O}(1)$ expansion parameters, and it is thus desirable to resum the contributions from multiple Glauber exchanges along with the SLLs. Whereas the resummation of the double-logarithmic series with only two Glauber exchanges has been understood recently~\cite{Becher:2021zkk,Becher:2023mtx}, we have shown in this article that, for quark-initiated scattering processes, also the contribution from additional Glauber exchanges can be resummed to all orders in the strong coupling $\alpha_s$ using renormalization-group techniques. To this end, we have generalized the appearing color traces by allowing for arbitrarily many insertions of the Glauber operator $\VG$ and of the soft-collinear, logarithmically-enhanced part of the anomalous dimension called $\Gammac$. Surprisingly, it turns out that the resulting color traces in \eqref{eq_color_trace_general} can be reduced to the same basis structures that appeared in the analysis of the series of SLLs. Working with a fixed coupling parameter $\alpha_s(\bar\mu)$, we have presented a formula for the resummation of the entire Glauber series \eqref{eq_Glauberseries_generic}, i.e.\ the leading double-logarithmic or $\pi^2$-enhanced quantum corrections for $2\to M$ scattering processes. Our main result is given in terms of multiple infinite sums, see \eqref{eq_rlsums} and \eqref{eq_master_formula}, but we have also presented closed analytic expressions in terms of simple trigonometric and polynomial functions in the expansion parameters $w$ and $w_\pi$ for the infinite sum of Glauber terms along with up to two insertions of $\Gammac$.

We have studied the numerical impact of the Glauber series for different partonic $2\to 2$, $2\to 1$, and $2\to 0$ scattering processes. In all cases it turns out that the contributions form higher-order Glauber terms (beyond the SLLs) are numerically small. While this was not to be expected from the numerical values of the expansion parameters $w$ and $w_\pi$ in \eqref{eq_wwpi}, we consider this finding as beneficial for future precision studies of non-global observables, since it will allow one to focus on the SLLs and subleading logarithmic corrections to them. We can attribute the suppression of the higher-order Glauber terms mainly to the smallness of the expansion coefficients $c_{\ell,n}$ with $\ell>1$ arising from the associated nested scale integrals, as well as to the alternating-sign behavior of the series. Indeed, quite generically we have observed a strong cancellation of relatively large terms in this alternating sum. 

In future work, it would be very important to generalize the exact results obtained here to the more general case, where the initial-state partons in the scattering process can be quarks or gluons. It would be exciting if also in this case an all-order solution could be obtained.

\pdfbookmark[1]{Acknowledgements}{Acknowledgements}
\subsubsection*{Acknowledgements}
We are grateful to Thomas Becher for many invaluable discussions. This work has been supported by the Cluster of Excellence Precision Physics, Fundamental Interactions, and Structure of Matter (PRISMA$^+$ EXC 2118/1) funded by the German Research Foundation (DFG) within the German Excellence Strategy (Project ID 39083149), and has received funding from the European Research Council (ERC) under the European Union’s Horizon 2022 Research and Innovation Programme (Grant agreement No.101097780, EFT4jets).

\appendix
\section{Resummed Glauber phases for \texorpdfstring{$\bm{n=2}$}{n=2}} 
\label{app_Glauber_phases_resummed_n_2}

Resumming infinitely many Glauber phases $\VG$ and simultaneously including $n=2$ insertions of $\Gammac$ leads to the following contribution to the $2\to M$ cross sections:
\begin{align}
	\label{eq_resummedVG_2}
	&\sum_{\ell=1}^{\infty} \, \sum_{\{r_i\},\, n = 2} \hat{\sigma}_{\{\underline{r}\}}^{\ell} 
	\nonumber \\
	&= \frac{\alpha_s(\bar{\mu}) L}{4\pi N_c} \, \frac{192w^2}{\kappa_0} \left(\frac{\sin \sqrt{x}}{x^{5/2}} - \frac{1}{x^2} + \frac{1}{6x} -\frac{1}{120}\right)
	\nonumber \\
	&\quad\times\left\{\sum_{j=3}^{2+M} J_j \sum_{i=2,4} c_i^{(2)} \tr{\H_{2\to M} \, \Oj_i} + J_{12} \sum_{i=5,6} d_i^{(2)} \tr{\H_{2\to M} \, \S_i}\right\}
	\nonumber \\
	&-\frac{\alpha_s(\bar{\mu}) L}{4\pi N_c} \, \frac{16w^2}{\kappa_0} \Bigg\{(x+5)\,\frac{\sin\sqrt{x}}{x^{5/2}}+3\,\frac{\cos\sqrt{x}}{x^2}-\frac{8}{x^2}+\frac{4}{3x}
	\nonumber \\
	&\quad+\frac{\kappa_1}{\kappa_0}\left((x+7)\,\frac{\sin\sqrt{x}}{x^{5/2}}+5\,\frac{\cos\sqrt{x}}{x^2}-\frac{12}{x^2}+\frac{8}{3x}-\frac{1}{10}\right)\Bigg\}
	\nonumber \\
	&\quad\times\left\{\sum_{j=3}^{2+M} J_j \sum_{i=2,4} c_i^{(1)} \tr{\H_{2\to M} \, \Oj_i} + J_{12} \sum_{i=5,6} d_i^{(1)} \tr{\H_{2\to M} \, \S_i}\right\}
	\nonumber \\
	&+\frac{\alpha_s(\bar{\mu}) L}{4\pi N_c} \, \frac{32w^2}{3\kappa_0} \Bigg\{3(2x-5)\,\frac{\sin\sqrt{x}}{x^{5/2}} - (x-15)\, \frac{\cos\sqrt{x}}{x^2}
	\nonumber \\
	&\quad+\frac{\kappa_2}{\kappa_0}\left(9(x-7)\,\frac{\sin\sqrt{x}}{x^{5/2}}-(x-33)\,\frac{\cos\sqrt{x}}{x^2}+\frac{30}{x^2}-\frac{2}{x}+\frac{3}{20}\right)\Bigg\}
	\nonumber \\
	&\quad \times \sum_{j=3}^{2+M} J_j \left(-\frac{1}{2}\,\tr{\H_{2\to M} \, \Oj_2} + \frac{1}{N_c} \, \tr{\H_{2\to M} \, \Oj_4}\right)
	\nonumber \\
	&+\frac{\alpha_s(\bar{\mu}) L}{4\pi N_c} \, \frac{2w^2}{\kappa_0} \Bigg\{(x^2-45x+105)\,\frac{\sin\sqrt{x}}{x^{5/2}}+5(2x-21)\,\frac{\cos\sqrt{x}}{x^2}
	\nonumber \\
	&\quad+\frac{1}{3}\frac{\kappa_1}{\kappa_0}\left(6(x^2-41x+175)\,\frac{\sin\sqrt{x}}{x^{5/2}}+2(28x-333)\,\frac{\cos\sqrt{x}}{x^2}-\frac{384}{x^2}+\frac{32}{x}\right)
	\nonumber \\
	&\quad+ \left(\frac{\kappa_1}{\kappa_0}\right)^2 \left((x^2-81x+441)\,\frac{\sin\sqrt{x}}{x^{5/2}}+(14x-249)\,\frac{\cos\sqrt{x}}{x^2}-\frac{192}{x^2}+\frac{16}{x}-\frac{4}{5}\right)\Bigg\}
	\nonumber \\
	&\quad \times \sum_{j=3}^{2+M} J_j \left(-\frac{1}{2}\,\tr{\H_{2\to M} \, \Oj_2} + \frac{1}{N_c} \, \tr{\H_{2\to M} \, \Oj_4}\right) ,
\end{align}
where $x = \kappa_0ww_{\pi}$ and $\kappa_r$ has been given in~\eqref{eq_kappar}.


\clearpage
\pdfbookmark[1]{References}{Refs}
\bibliography{refs.bib}

\end{document}